\documentclass[prc,preprint,
  showpacs,showkeys,lengthcheck,
  nofootinbib,tightenlines,onecolumn,notitlepage,
  preprintnumbers,superscriptaddress]{revtex4-1}

\usepackage{graphicx}
\usepackage[usenames,dvipsnames]{xcolor}
\graphicspath{{figures/}}

\usepackage{hyperref}
\hypersetup{colorlinks=true,citecolor=blue,linkcolor=blue,urlcolor=blue}

\usepackage{diagbox}
\usepackage{booktabs}

\usepackage{amsmath,amssymb,amsfonts}
\usepackage{mathrsfs}
\usepackage{slashed}
\usepackage{bm}

\def\g{\gamma}\def\p{\partial}\def\m{\mu}\def\ph{\phi}
\newcommand{\eq}[1]{Eq.~(\ref{#1})}
\def\tz{{\tilde z}}
\def\G{\Gamma}\def\b{\beta}
\def\s2{\sqrt{2}}

\begin{document}

\title{Mass decomposition of the pion in the 't Hooft model}

\author{Adam Freese}
\email{afreese@uw.edu}
\address{Department of Physics, University of Washington, Seattle, WA 98195, USA}

\author{Gerald A. Miller}
\email{miller@uw.edu}
\address{Department of Physics, University of Washington, Seattle, WA 98195, USA}

\begin{abstract}
  We obtain the energy-momentum tensor (EMT) in the 't Hooft model of
  two-dimensional quantum chromodynamics.
  The EMT is decomposed into contributions from quark and gluon fields,
  with all of the (plus component of) the light front momentum
  being carried by the quark field.
  The energy is split between quark and gluon fields,
  with the gluon field carrying the self-energy of the dressed quarks.
  We consider the pion in the limit of small but non-zero quark masses---which
  has previously withstood numerical treatment---as
  as a concrete example.
  We solve for the pion wave function using a variational method
  and obtain numerical results for its energy breakdown into
  quark and gluon contributions.
\end{abstract}

\preprint{NT@UW-22-16}

\maketitle


\section{Introduction}

Recently there has been an increasing effort to understand the origin of the
proton's mass in terms of QCD; see for instance
Refs.~\cite{Ji:1994av,He:1994gz,Metz:2020vxd,Lorce:2021xku,Ji:2021qgo}.
Since the mass plays multiple vital roles within a quantum field
theory---such as a Lorentz scalar, a rest frame energy, and a central
charge of the light front's Galilean subgroup---there are consequently
many approaches one can take to understanding the mass~\cite{Lorce:2021xku}.

One of the most common approaches is to analyze and decompose the mass
through the trace of the energy momentum tensor (EMT):
\begin{align}
  2M^2
  =
  \langle p | T^{\mu}_{\phantom{\mu}\mu}(0) | p \rangle
  \,,
\end{align}
see for instance Refs.~\cite{Crewther:1972kn,Chanowitz:1972vd,Collins:1976yq,Shifman:1978zn}.
Another common approach, which provides different physical insights,
is to analyze the mass as the rest-frame energy associated with a Hamiltonian density,
and to decompose the Hamiltonian into contributions arising
from the quark and gluon fields; see for instance Refs.~\cite{Ji:1994av,Metz:2020vxd,Lorce:2021xku,Ji:2021qgo}.
The latter approach depends on the form of relativistic dynamics used~\cite{Dirac:1949cp},
since for instance the instant form energy density is given by $T^{00}(x)$
and the light front energy density by $T^{+-}(x)$.

A case of special interest occurs for quantum chromodynamics (QCD) in $(1+1)$ dimensions.
In this case, the light front Hamiltonian density and trace of the EMT are
both proportional to $T^{+-}$,
meaning that both approaches will produce the same mass decomposition.
Moreover, $(1+1)$-dimensional QCD is UV finite, and subtleties about
renormalization and scheme dependence of the decomposition
(which are at the heart of a controversy about the proton mass decomposition
in $(3+1)$ dimensions~\cite{Lorce:2021xku}) are avoided.
This makes $(1+1)$-dimensional QCD a promising avenue to explore
a mass decomposition within QCD itself.

Besides the proton, understanding how QCD dynamics gives rise to a
nearly massless pion is a question of special concern,
since the pion plays a special role as the Nambu-Goldstone boson
of chiral symmetry breaking~\cite{Nambu:1961tp,Nambu:1961fr,Gell-Mann:1968hlm,Pagels:1978ba,Chang:2013pq}.
A fully satisfactory understanding of proton mass decomposition
cannot be had without an understanding of the pion mass decomposition as well.

The 't Hooft model~\cite{tHooft:1974pnl} is an especially promising avenue for approaching
the mass decomposition of the pion.
This model is defined as QCD in $(1+1)$ dimensions in the limit that
$N_c\rightarrow\infty$ but $g^2 N_c$ is held fixed.
The 't Hooft model was extensively studied during the 1970's
and has had much success in describing qualitative properties of mesons,
such as confinement and Regge trajectories
(see the review~\cite{Ellis:1977mw}).

Moreover, there has been a recent revival of interest in the 't Hooft model and other treatments of $(1+1)$-dimensional
QCD~\cite{DeTeramond:2021jnn,Li:2021jqb,Ahmady:2021lsh,Ahmady:2021yzh}
stemming from the need to include the effects of non-vanishing quark masses
and to enlarge the number of space-time variables in light-front holographic QCD to four.
In the original treatments of light front holography (see the review~\cite{Brodsky:2014yha})
the chiral limit is used and the longitudinal light-front momentum fraction $x$ is frozen~\cite{Sheckler:2020fbt},
so that effectively the only degrees of freedom are light-front time and transverse position.
The first effort aimed at including the effects of mass was contained in Ref.~\cite{Chabysheva:2012fe}.
There has been much numerical work, but the case of small but non-zero quark masses
has sustained numerical challenges~\cite{PhysRevD.19.3024}.
Thus, the case of small but non-zero quark mass is also of special interest.

In this work, we examine the energy-momentum tensor in the 't Hooft model
in the low-mass domain,
obtaining a mass decomposition for the pion into quark and gluon contributions.
We define the model in Sec.~\ref{sec:model},
and obtain and examine the EMT operator in Sec.~\ref{sec:emt}.
In Sec.~\ref{sec:pion} we numerically consider the pion as a case of interest,
solving for the wave function through a variational method
and obtaining numerical results for the mass decomposition.
The results of this method are found to be valid specifically in the domain
of small quark masses.
We conclude in Sec.~\ref{sec:end}.


\section{Model and definitions}
\label{sec:model}

The 't Hooft model is $D=(1+1)$-dimensional QCD in the large $N_c$ limit
(but specifically with $g^2 N_c$ held fixed)~\cite{tHooft:1974pnl}.
The Lagrangian is the standard QCD Lagrangian:
\begin{align}
  \mathscr{L}
  =
  \bar{q}(x)
  \left( \frac{i}{2} \overleftrightarrow{\slashed{D}} - m_q \right)
  q(x)
  -
  \frac{1}{4} F_{\mu\nu}^a(x) F_a^{\mu\nu}(x)
  \,,
\end{align}
where $q(x)$ is a column vector of quarks of different flavors
and $m_q$ is a mass matrix.
The covariant derivative (in the defining rep) is
\begin{align}
  D_\mu q(x)
  =
  \partial_\mu q(x)
  -
  ig A^a_\mu(x) T_a q(x)
  \,,
\end{align}
where $T_a$ is a generator of $\mathrm{SU}(N_c,\mathbb{C})$.
To make calculations easier, the  light cone gauge is used:
\begin{align}
  A_a^+(x) &= 0
  \,.
\end{align}
In this case the only non-zero components of
gluon field strength tensor are given by:
\begin{align}
  F_a^{+-}(x)
  =
  -
  F_a^{-+}(x)
  =
  \partial_- A_a^-(x)
  \,.
\end{align}
An immediate consequence of using light cone gauge
is that the Lagrangian can be written:
\begin{align}
  \mathscr{L}
  =
  \bar{q}(x)
  \left( \frac{i}{2} \overleftrightarrow{\slashed{\partial}} - m_q \right)
  q(x)
  +
  g
  \bar{q}(x)
  A_a^-(x)
  \gamma^+
  T_a
  q(x)
  +
  \frac{1}{2} \big(\partial_- A_a^-(x)\big)^2
  \,.
\end{align}


\subsection{Dynamical degrees of freedom}

It is useful to separate the quark field $q(x)$ into
independent $q_+$ and dependent $q_-$ terms using the projection operators
$\Pi^\pm\equiv {1\over2}\g^\mp\g^\pm$ and $q_\pm=\Pi^\pm q$,
with $\g^\pm ={1\over\sqrt{2}}(\g^0\pm\g^3)$.
Then the Lagrangian can be rewritten as:
\begin{align}
  \mathscr{L}
  =
  \s2 {q}_+^\dagger i\p^-q_+
  + \s2 {q}_-^\dagger i\p^+q_-
  -{m_q\over\s2}
  \Big( q_+^\dagger\g^-q_- + q_-^\dagger\g^+q_+ \Big)
  +
  g
  \s2{q}_+^\dagger
  A_a^-
  T_a
  q_+
  +
  \frac{1}{2} \big(\partial_- A_a^-\big)^2
  \,,
\end{align}
where dependence on $x$ was suppressed to compactify the formula.
The fields $q_+$ and $A^-_a$
are not dynamical fields because their time derivative does not appear in the Lagrangian.
This means that they can be rewritten in terms of the independent field operator $q_+$.
The Euler-Lagrange equation for $A_a^-$ is given by
\begin{align}
  \partial_-^2 A_a^-(x)
  =
  g\bar{q}(x) \gamma^+ T_a q(x)
  \,,
\end{align}
the general solution being:
\begin{align}
  \label{eqn:A-}
  A_a^-(x)
  =
  \frac{g}{2}
  \sum_q
  \int_{-\infty}^\infty \mathrm{d}z \,
  |x^--z|
  \bar{q}(z) \gamma^+ T_a q(z)
  +
  x^- B_a(x^+)
  +
  C_a(x^+)
  \,.
\end{align}
The function $C_a(x^+)$ can be removed by making a gauge transformation.
The term $B_a(x^+)$
is set to zero so as to impose the boundary condition that $A_-^a(x)=0$ in the absence of quark sources.

The Euler-Lagrange equation for the quark field is:
\begin{align}\label{deq}
  \Big(
  i \gamma^+ \partial_+
  +
  i \gamma^- \partial_-
  +
  g T_a A_a^-(x) \gamma^+
  -
  m_q
  \Big) q(x)
  =
  0
  \,.
\end{align}
Multiplying \eq{deq} by $\g^+$ leads to the results:
\begin{align}
  q_-
  &=
  {m_q\g^0\over\sqrt{2}i\partial^+}q_+
  \\
  \Big(i\p^++gT_aA^-_a\Big)q_+
  &=
  m{\g^0\over\sqrt{2}}q_-={m_q^2\over 2i\partial^+}q_+
  \,,
\end{align}
where the inverse operator is defined:
\begin{align}
  \frac{1}{\partial^+}
  f(x^+,x^-)
  =
  \frac{1}{2}
  \int_{-\infty}^\infty \mathrm{d}z^- \,
  \epsilon(x^--z^-)
  f(x^+,z^-)
  \,.
\end{align}
With these results, the solution for the gluon field is:
\begin{align}
  A_a^-
  =
  \sqrt{2}g
  \frac{1}{(\partial^+)^2}
  q_+^\dagger T_a q_+
  \,.
\end{align}


\section{Energy-momentum tensor}
\label{sec:emt}

Let us consider components of the energy-momentum tensor,
for which we use the standard symmetric, Belinfante EMT of QCD~\cite{Kugo:1979gm,Leader:2013jra}.
In light cone coordinates, $P^+$ is a kinematic momentum and $P^-$ is the
Hamiltonian, and accordingly $T^{++}$ is a momentum density and $T^{+-}$
is a Hamiltonian (energy) density.
It is conventional in much of the hadron physics literature
(see e.g.\ Refs.~\cite{Ji:1994av,Leader:2013jra,Lorce:2021xku})
to decompose the EMT into gauge-invariant ``quark'' and ``gluon'' contributions:
\begin{subequations}
  \label{eqn:Pqg}
  \begin{align}
    T^{\mu\nu}
    &=
    T_q^{\mu\nu}
    +
    T_g^{\mu\nu}
    \\
    T_q^{\mu\nu}
    &=
    \frac{i}{4}\Big(
    \bar{q}(x) \gamma^\mu \overleftrightarrow{\partial}^\nu q(x)
    +
    \bar{q}(x) \gamma^\nu \overleftrightarrow{\partial}^\mu q(x)
    \Big)
    +
    \frac{g}{2}
    \Big(
    \bar{q}(x) \gamma^\mu A^\nu(x) q(x)
    +
    \bar{q}(x) \gamma^\nu A^\mu(x) q(x)
    \Big)
    \\
    T_g^{\mu\nu}
    &=
    g_{\rho\sigma} F_a^{\mu\rho}(x) F_{a}^{\sigma\nu}(x)
    +
    \frac{g^{\mu\nu}}{4}
    \big(F_a^{\rho\sigma}(x)\big)^2
    \,.
  \end{align}
\end{subequations}
Notably, interaction terms containing both the quark and gluon fields
are present in the ``quark'' piece of the EMT.
This is done in order to ensure gauge invariance of the breakdown.

In light cone gauge, the gluon contribution to the momentum $P^+$ is zero,
and the quark contribution is:
\begin{align}
  P^+
  =
  P^+_q
  =
  \int_{-\infty}^\infty \mathrm{d}x^- \,
  {\sqrt{2}}q_+^\dagger i\p^+q_+
  \,.
\end{align}
The quark and gluon contributions to the energy $P^-$
can be written in terms of the independent $q_+$ fields as:
\begin{subequations}
  \begin{align}
    P^-_q
    &=
    \int_{-\infty}^\infty \mathrm{d}x^- \,
    q_+^\dagger {m_q^2 \over\sqrt{2} i\partial^+}q_+
    \\
    P^-_g
    &=
    g^2
    \int_{-\infty}^\infty \mathrm{d}x^- \,
    \left(
    \frac{1}{\partial^+} (q_+^\dagger T_a q_+)
    \right)^2
    \,.
  \end{align}
\end{subequations}
Summing over the quark and gluon pieces reproduces the finding of
Ehlers~\cite{Ehlers:2022oal}:
\begin{align}
  P^-
  =
  P_q^-
  +
  P_g^-
  =
  \int_{-\infty}^\infty \mathrm{d}x^-
  \left\{
    q_+^\dagger {m_q^2 \over\sqrt{2} i\p^+}q_+
    +
    g^2\left({1\over\p^+}(q_+^\dagger T^a q_+)\right)^2
    \right\}
  \,.
\end{align}
In addition to the breakdown of the operator $P^-$ itself,
it is also worth looking at form factors.
At zero momentum transfer, the separate quark and gluon contributions
to the EMT are parametrized by two quantities each:
\begin{align}
  \langle p | T^{\mu\nu}_{a}(0) | p \rangle
  =
  2 p^\mu p^\nu A_a(0)
  +
  2 g^{\mu\nu} \mathcal{M}^2 \bar{c}_a(0)
  \,,
\end{align}
where $a=q,g$ and where $\mathcal{M}$ is the mass of the hadron.
The quantities satisfy the sum rules:
\begin{subequations}
  \begin{align}
    A_q(0) + A_g(0) &= 1
    \\
    \bar{c}_q(0) + \bar{c}_g(0) &= 0
    \,.
  \end{align}
\end{subequations}
Since $T^{++}_g = 0$ in the 't Hooft model,
we have $A_q(0) = 1$ and $A_g(0) = 0$.
Moreover, from integrating the $+-$ components, we have:
\begin{subequations}
  \label{eqn:cbar}
  \begin{align}
    2 p^+ P_q^-
    &=
    \mathcal{M}^2 \Big( 1 + 2 \bar{c}_q(0) \Big)
    \\
    2 p^+ P_g^-
    &=
    2 \mathcal{M}^2 \bar{c}_g(0)
    =
    - 2 \mathcal{M}^2 \bar{c}_q(0)
    \,.
  \end{align}
\end{subequations}


\subsection{Effective potential and quark dressing}

\begin{figure}
  \includegraphics{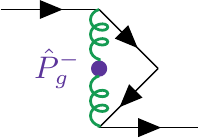}
  ~~~~
  \includegraphics{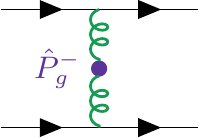}
  ~~~~
  \includegraphics{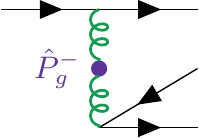}
  ~~~~
  \includegraphics{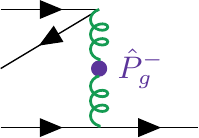}
  ~~~~
  \includegraphics{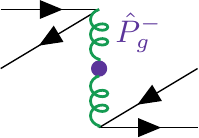}
  \caption{
    Diagrammatic representation of the possible scenarios encoded in
    the gluon energy operator $P_g^-$.
    The first diagram is a quark-self energy which is absorbed into the
    effective kinetic energy (first term) of Eq.~(\ref{eqn:pm}).
    The remaining diagrams are contained in the normal-ordered effective
    potential of Eq.~(\ref{eqn:pm}).
    In the large $N_c$ limit, all diagrams except the first two are suppressed.
  }
  \label{fig:ordering}
\end{figure}

It is reasonable to interpret $P_q^-$ as a kinetic energy for the dynamical field $q_+$
and $P_g^-$ as a potential energy,
as remarked by Ehlers~\cite{Ehlers:2022oal}.
This gluon energy (potential energy) can be rewritten as a non-local operator using the inverse operator:
\begin{subequations}
  \begin{align}
    P_g^-
    &=
    \int_{-\infty}^\infty \mathrm{d}x^- \,
    \int_{-\infty}^\infty \mathrm{d}y^- \,
    V(x^-,y^-)
    \\
    V(x^-,y^-)
    & =
    -{g^2\over2}
    |x^--y^-|J^a(x^-) J^a(y^-)
    \\
    J^a(x^-)
    &=
    q_+^\dagger(x^-) T^a q_+(x^-)
    \,.
  \end{align}
\end{subequations}
The effective interaction $V$ can further be broken down into
a normal ordered operator that encodes a static quark-antiquark potential,
and a single quark operator that encodes a quark self-energy (see Fig.~\ref{fig:ordering}).
This is done by using the fundamental commutation relation between $q_+^\dagger $ and $q_+$,
and is described in depth by Ehlers~\cite{Ehlers:2022oal}.
The quark self-energy term can combined with the quark mass $m_q$ to produce
a dressed mass $M_q$, given in the large $N_c$ limit as~\cite{tHooft:1974pnl}:
\begin{align}
  M_q^2
  =
  m_q^2
  -
  g^2N_c/\pi
  \,.
\end{align}
When $m_q^2 < g^2 N_c/\pi$, the dressed mass $M_q$ will be imaginary.
A complex dressed mass indicates that confinement occurs,
as happens to gluons for instance in
the Gribov theory of gluon confinement~\cite{Gribov:1977wm}.

In terms of the dressed mass and static potential, the Hamiltonian can be written:
\begin{align}
  P^-
  &=
  \int_{-\infty}^\infty \mathrm{d}x^- \,  q_+^\dagger {M_q^2 \over\sqrt{2} i\p^+}q_+
  +
  \int_{-\infty}^\infty \mathrm{d}x^-
  \int_{-\infty}^\infty \mathrm{d}y^- \,
  :\! V(x^-,y^-)\! :
  \,.
  \label{eqn:pm}
\end{align}
The first term in this expression mixes contributions from $P^-_q$ and $P^-_g$.
This occurs because the dressed quark mass contains contributions from the gluon field,
as depicted in Fig.~\ref{fig:ordering}.
The breakdown suggested by this expression presents a reasonable alternative decomposition
within the context of the model:
an \emph{effective} kinetic energy carried by dressed quarks
and an \emph{effective} static potential between them.
A curious aspect of this is that the effective kinetic energy can be negative if $M_q^2 < 0$.

In this work, we prioritize the breakdown in Eq.~(\ref{eqn:Pqg}),
since it attributes energy contributions to the bare quark and gluon fields
in a gauge-invariant way.


\subsection{Application to mesons}

A meson state consisting of a quark $q_1$ and antiquark $\bar{q}_2$
with fixed momentum $p^+$ can be written in terms of creation
and annihilation operators as~\cite{Ehlers:2022oal}:
\begin{align}
  | p^+ \rangle
  =
  \sum_{c=1}^{N_c}
  \int_0^1 \mathrm{d}x \,
  \frac{ \phi(x) }{ \sqrt{4\pi N_c x(1-x)} }
  b_{c,1}^\dagger(xp^+)
  d_{c,2}^\dagger((1-x)p^+)
  | 0 \rangle
  \,,
\end{align}
where $x$ is the fraction of the meson's momentum carried by the quark.
This meson ket satisfies the normalization rule
\begin{align}
  \langle p'^+ | p^+ \rangle
  =
  4\pi p^+ \delta(p^+ - p'^+)
\end{align}
provided that:
\begin{align}
  \label{eqn:norm}
  \int_0^1 \mathrm{d}x \,
  |\phi(x)|^2
  =
  1
  \,.
\end{align}
The function $\phi(x)$ is thus interpreted as the meson wave function
over the momentum fraction $x$.
Acting on $|p^+\rangle$ with the Hamiltonian as expressed in Eq.~(\ref{eqn:pm})
gives the 't Hooft equation:
\begin{align}
  \mathcal{M}^2
  \phi(x)
  =
  \left({M_1^2\over x}+{M_2^2\over 1-x}\right)
  \phi(x)
  +
  \frac{g^2N_c}{\pi} P\int_0^1\mathrm{d}y \, {1\over (x-y)^2}
  \phi(y)
  \label{th1}
  \,,
\end{align}
where ${\cal M}^2$ is the eigenvalue of the operator $2p^+P^-$.
The principal value is defined according to~\cite{tHooft:1974pnl} as
\begin{align}
  P {f(x,y)\over (x-y)^2}
  \equiv
  {1\over 2}\left[ {f(x,y)\over (x-y +i \epsilon)^2}+{f(x,y)\over (x-y -i \epsilon)^2}\right]
  \label{eqn:pv}
  \,,
\end{align}
where the limit $\epsilon\to0$ is to be taken after integration.
Expanding out the dressed masses in terms of the bare masses and dressing
gives an alternative form of the 't Hooft equation:
\begin{align}
  \label{eqn:thooft:general}
  \mathcal{M}^2
  \phi(x)
  =
  \left({m_1^2\over x}+{m_2^2\over 1-x}\right)
  \phi(x)
  -
  \frac{g^2N_c}{\pi} P\int_0^1\mathrm{d}y \,
  \frac{\phi(x) - \phi(y)}{(x-y)^2}
  \,.
\end{align}
In this form, the first term corresponds to the quark energy
as $2p^+ P_q^-$ and the second to the gluon energy as $2p^+ P_g^-$.


\section{Pions in the low-mass limit}
\label{sec:pion}

Let us consider the pion in the limit of small but non-zero quark masses,
a case that has until now presented numerical challenges~\cite{PhysRevD.19.3024}.

For equal current mass $m$ quarks, the 't Hooft equation in Eq.~(\ref{eqn:thooft:general})
simplifies to:
\begin{align}
  \label{eqn:thooft}
  \mu^2
  \phi(x)
  =
  \frac{\gamma}{x(1-x)}
  \phi(x)
  -
  \int_0^1\mathrm{d}y \,
  \frac{\phi(x) - \phi(y)}{(x-y)^2}
  \,,
\end{align}
where $\m^2\equiv \frac{{\cal M}^2\pi}{g^2N_c}$ and $\g\equiv{ \pi m^2\over g^2 N_c}$
are unitless quantities to simplify the formula.
As discussed in Sec.~\ref{sec:emt}, the first and second terms on the right-hand side
correspond to the quark energy $P^-_q$ and the gluon energy $P_g^-$, respectively.
The equation can also alternatively be written in the more standard form:
\begin{align}
  \mu^2
  \phi(x)
  =
  \frac{\gamma-1}{x(1-x)}
  \phi(x)
  +
  \int_0^1\mathrm{d}y \,
  \frac{\phi(y)}{(x-y)^2}
  \,,
\end{align}
where the first and second terms now correspond to a dressed quark effective kinetic energy
and a static confining potential.

't Hooft postulated the ansatz:
\begin{align}
  \label{eqn:ansatz}
  \phi(x)=x^\b(1-x)^\b\frac{\sqrt{\Gamma (4 \beta +2)}}{\Gamma (2 \beta +1)}
  \,,
\end{align}
to be valid at the end-points and to cancel the end-point singularities appearing in the effective kinetic energy term.
The multiplicative factor ensures that Eq.~(\ref{eqn:norm}) is obeyed.
The integral over $y$ in Eq.~(\ref{eqn:thooft}) can be approximated as
$x^{\b-1}[\pi \b \cot\pi \b-1]$ for very small values of $x$,
so expanding Eq.~(\ref{eqn:thooft}) for very small values of $x$ leads to:
\begin{align}
  {\pi m^2\over g^2 N_c}-1 +\pi \b \cot\pi \b
  =
  0
  \,.
\end{align}
If $m=0$ this equation is solved with $\b=0$,
and continuity ensures that small values of $m$ entail small values of $\beta$.
Expanding around $\beta=0$ gives the approximate solution for small quark masses of:
\begin{align}
  \label{eqn:beta}
  \b
  \approx
  \sqrt{3\over \pi N_c }{m\over g}
  \,.
\end{align}
With this value of $\beta$, we can also derive a relationship
between the pion mass and current quark mass,
with the pion defined as the ground state of the quark-antiquark system.
Integrating Eq.~(\ref{eqn:thooft}) over $x$ gives:
\begin{align}
  \m^2
  =
  \frac{
    \gamma
    \int_0^1 \mathrm{d}x\, {\ph(x)\over x(1-x)}
  }{
    \int_0^1 \mathrm{d}x\, \ph(x)
  }
  \,,
\end{align}
and evaluating this with the ansatz of Eq.~(\ref{eqn:ansatz})
along with the found value of $\beta$ gives:
\begin{align}
  \label{eqn:gmor}
  m_\pi^2
  \approx
  2g\sqrt{\pi N_c\over 3} m
  +
  4m^2
  \,,
\end{align}
reminiscent of the Gell-Mann--Oakes--Renner (GMOR) relation~\cite{Gell-Mann:1968hlm}.
It's worth noting that this result was obtained using the ansatz of
Eq.~(\ref{eqn:ansatz}) for the wave function at \emph{all} values of $x$,
despite the form originally being postulated only for $x$ near the endpoints.
It is therefore important to verify the validity of the wave function
Eq.~(\ref{eqn:ansatz}) at all $x$, at least as an approximate form,
as well as of Eq.~(\ref{eqn:gmor}) within the 't Hooft model.

To proceed, we take a variational approach in which Eq.~(\ref{eqn:ansatz}) is used
as a trial wave function, and the value of $\beta$ is determined by minimizing the expectation
value of $\mu^2(\beta)$:
\begin{align}
  \label{eqn:variation}
  \m^2(\b)
  =
  (\g-1)
  \int_0^1 \mathrm{d}x \,  \frac{|\phi(x)|^2}{x(1-x)}
  +
  \int_0^1 \mathrm{d}x \,
  \int_0^1 \mathrm{d}y \,
  \ph(x) {P\over (x-y)^2} \phi(y)
  \,.
\end{align}
Using Eq.~(\ref{eqn:ansatz}) we find that:
\begin{align}
  \int_0^1 \mathrm{d}x\,
  {\ph^2(x)\over x(1-x)}
  =
  4+{1\over \b}
  \,.
\end{align}
The double integral appearing in Eq.~(\ref{eqn:variation}),
which corresponds to the effective static potential between dressed quarks,
can be evaluated in closed form by going to the coordinate-space representation:
\begin{align}
  V_{\mathrm{eff}}(\b)
  =
  \int_0^1 \mathrm{d}x \,
  \int_0^1 \mathrm{d}y \,
  \ph(x) {P\over (x-y)^2} \phi(y)
  =
  {1\over 2}
  \int_{-\infty}^\infty \mathrm{d}\tz \,
  \phi^*(\tz) |\tilde z|\tilde\phi(\tz)
  \,,
\end{align}
where:
\begin{align}
  \tilde\ph(\tz)
  =
  \int_0^1 \mathrm{d}y \,
  \ph(y) e^{i y \tz}
  \label{eqn:phiz}
  \,.
\end{align}
Direct evaluation gives:
\begin{align}
  \ph(\tz)
  =
  e^{i\tz/2}\sqrt{\pi } 2^{1-2 (\beta +1)} \Gamma (\beta +1) \, _0\tilde{F}_1\left(\beta +\frac{3}{2};-\frac{z^2}{16}\right)\frac{\sqrt{\Gamma (4 \beta +2)}}{\Gamma (2 \beta +1)}
  \,,
\end{align}
where $_0\tilde{F}_1$ is a regularized confluent hypergeometric function~\cite{NIST:DLMF}.
It can be shown that:
\begin{align}
  \int_{-\infty}^\infty \mathrm{d}z \,
  |z|\,_0\tilde{F}_1^2\left(\beta +\frac{3}{2};-\frac{z^2}{16}\right)={8\over \b\, \G^2(1/2+\b)}
  \,,
\end{align}
and using this gives, after simplification:
\begin{align}
  V_{\mathrm{eff}}(\b)
  =
  \pi^2\,2^{-8\b} {\G(2+4\b)\over \b\G^4(1/2+\b)}
  \,.
\end{align}
The expectation value of the Hamiltonian is thus:
\begin{align}
  \mu^2(\beta)
  =
  (\gamma-1)\left(4+{1\over \b}\right)
  +
  \pi^2\,2^{-8\b} {\G(2+4\b)\over \b\G^4(1/2+\b)}
  \,.
\end{align}
A minimum exists for this function, since its asymptotic forms
at small and large $\beta$ are:
\begin{subequations}
  \begin{align}
    \lim_{\b\to0} \mu^2(\b)
    &=
    \frac{\gamma}{\beta}
    \\
    \lim_{\b\to\infty} \mu^2(\b)
    &=
    2 \sqrt{2 \pi } \sqrt{\beta }
    \,,
  \end{align}
\end{subequations}
meaning a minimum (and thus a solution) exists.

\begin{figure}
  \includegraphics[width=0.49\textwidth]{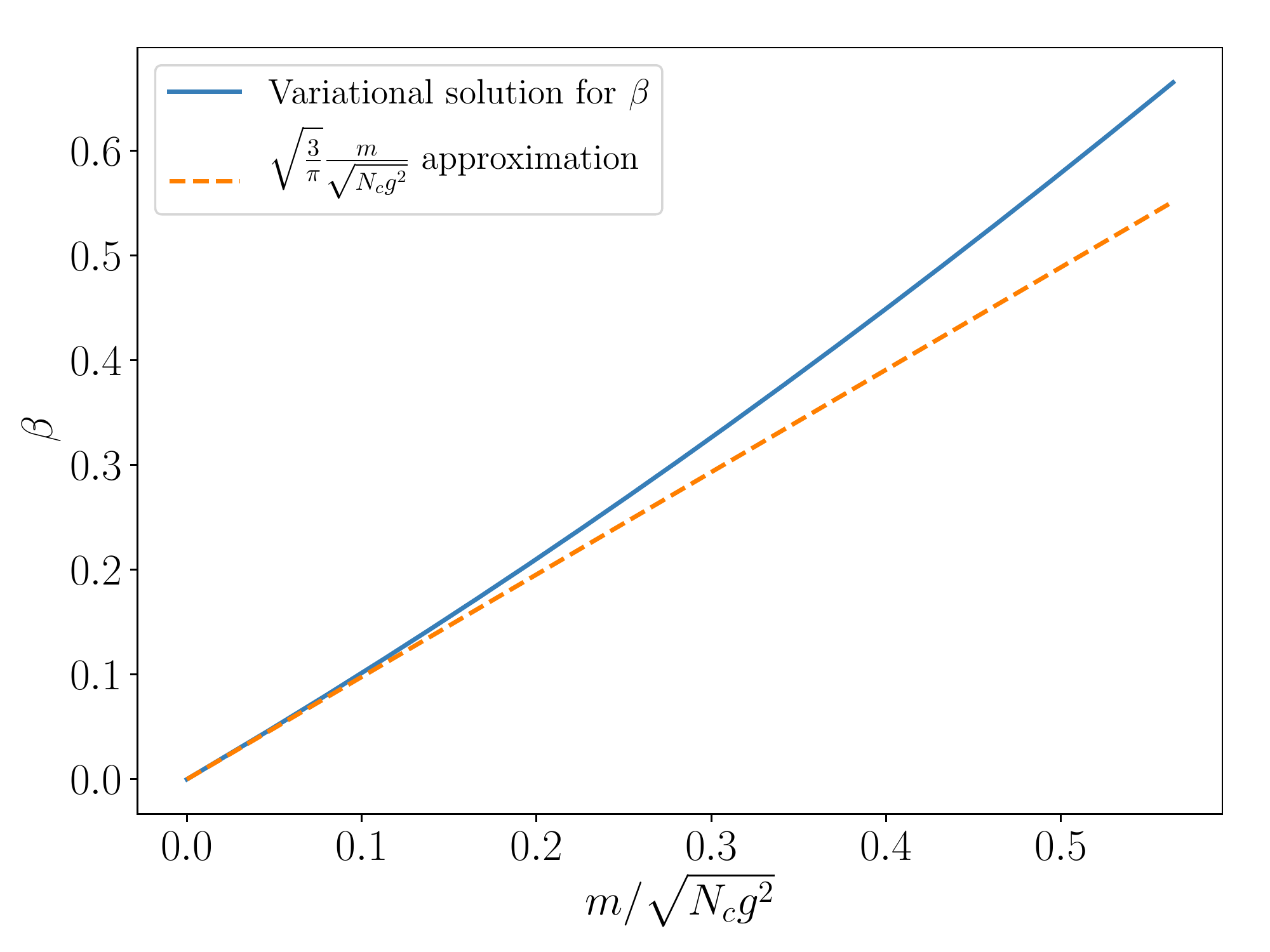}
  \includegraphics[width=0.49\textwidth]{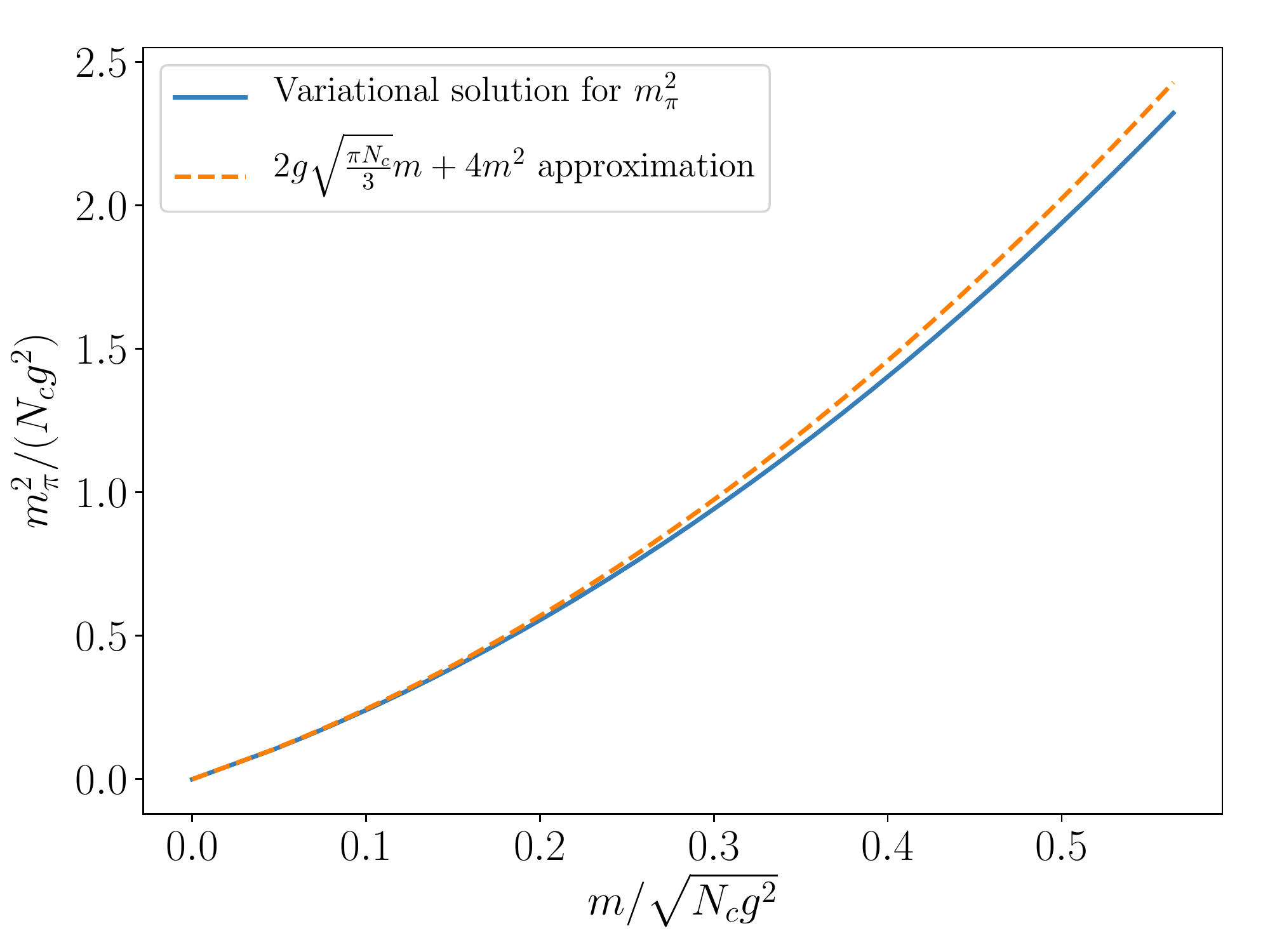}
  \caption{
    Numerical results for $\beta$ obtained from minimizing
    the Hamiltonian associated with the ansatz of Eq.~(\ref{eqn:ansatz}).
    (Left panel) gives the results for $\beta$,
    comparing to the approximation of Eq.~(\ref{eqn:beta}).
    (Right panel) gives the results for $m_\pi^2$,
    comparing them to the approximate result of Eq.~(\ref{eqn:gmor}).
  }
  \label{fig:results}
\end{figure}

The minimum of $\mu^2(\beta)$ has been determined numerically
for several values of the quark mass $m$, with results shown in Fig.~\ref{fig:results}.
The left panel shows that Eq.~(\ref{eqn:beta}) holds nearly exactly for quark masses
as large as around $0.1$ in units of $\sqrt{N_c g^2}$,
and for these values the variational solution can be reasonably be considered nearly exact.
For larger quark masses, the ansatz of Eq.~(\ref{eqn:ansatz}) becomes less reliable,
but perhaps useful nonetheless as a rough approximation.
The right panel, interestingly, shows that the GMOR-like relation of
Eq.~(\ref{eqn:gmor}) holds to good precision well past $m/\sqrt{N_c g^2} \sim 0.1$.

\begin{figure}
  \includegraphics[width=0.49\textwidth]{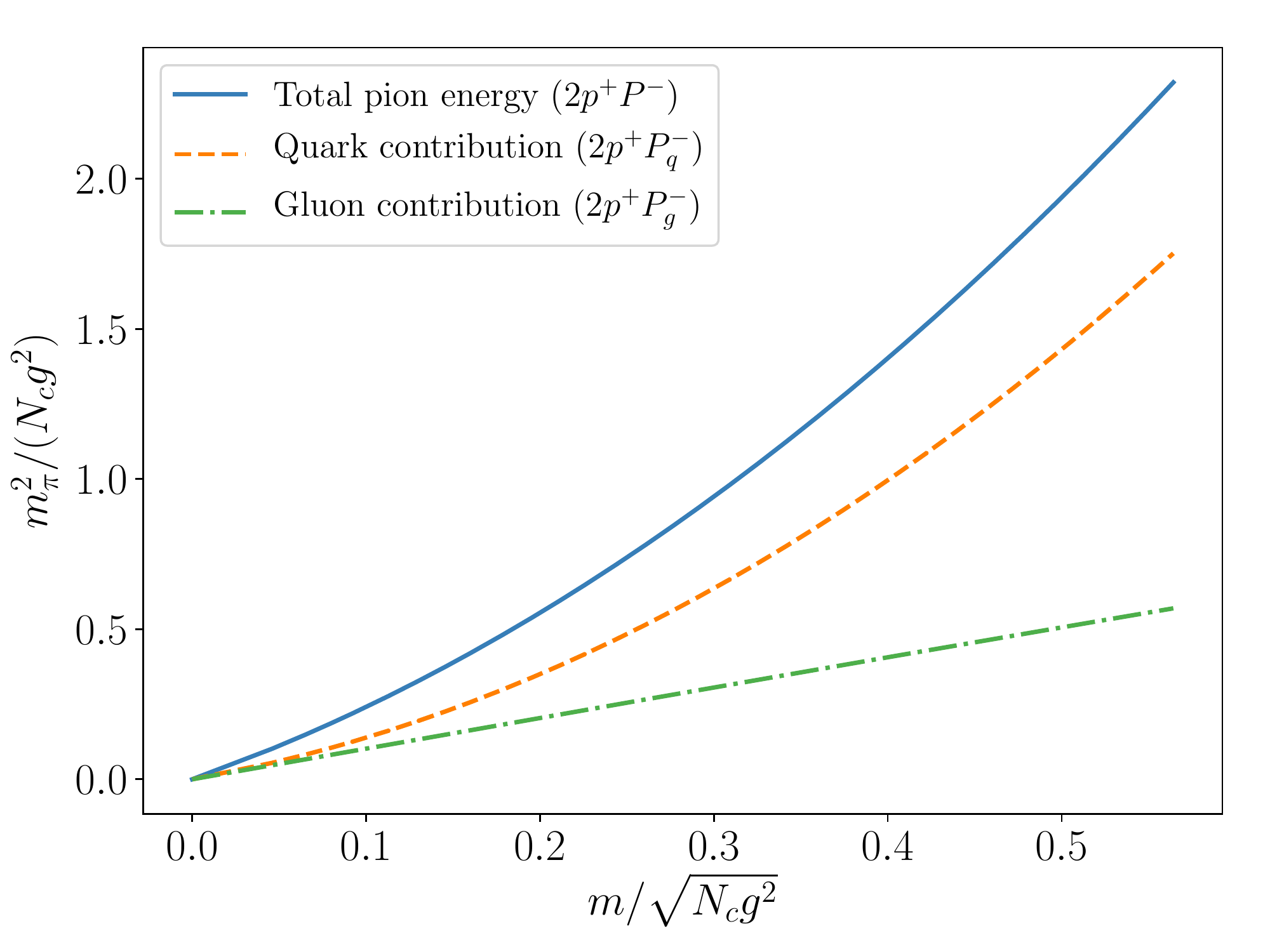}
  \includegraphics[width=0.49\textwidth]{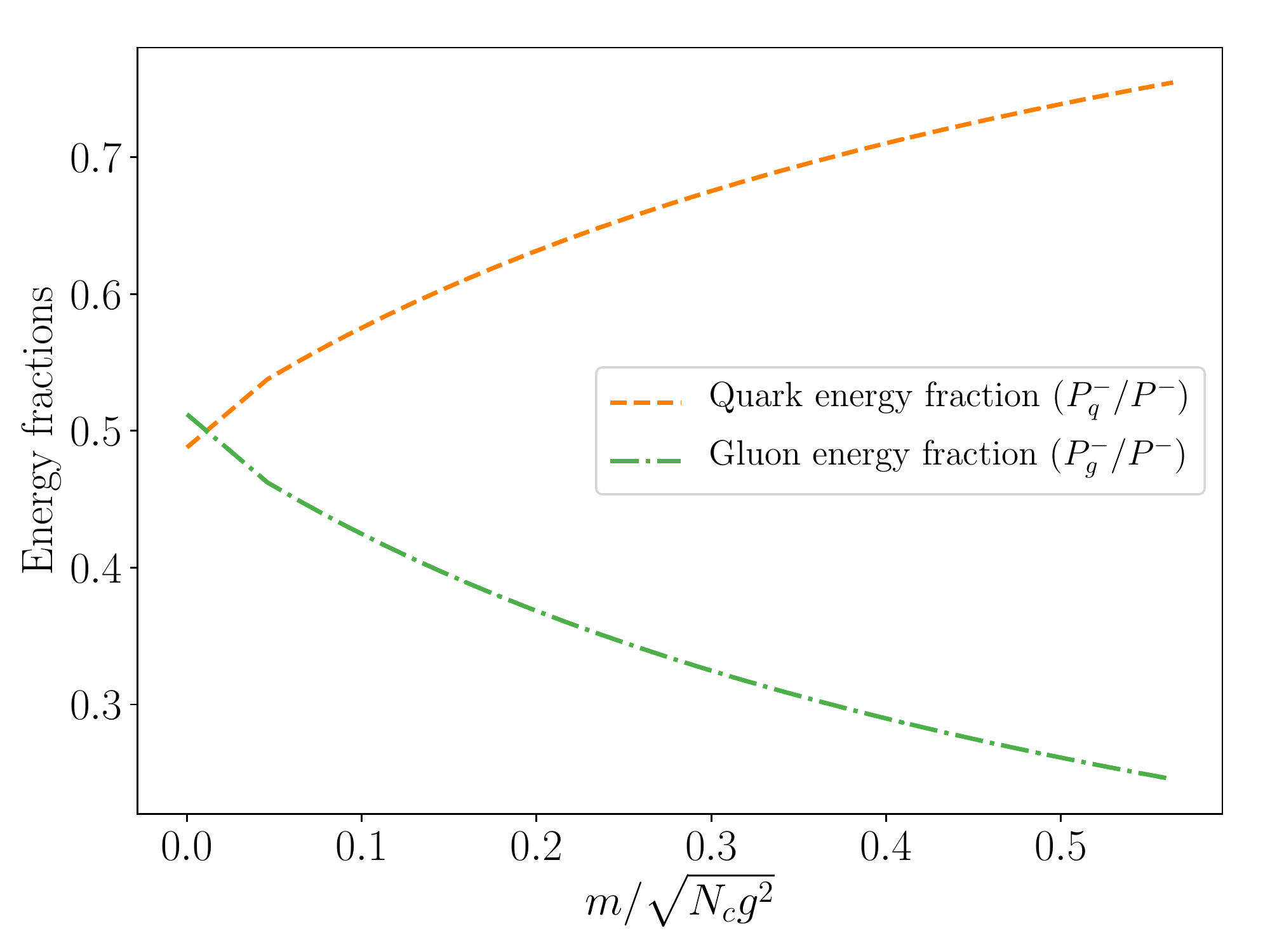}
  \caption{
    Breakdown of the pion energy in the variational solution to the 't Hooft model.
    (Left panel) compares the total pion energy $2p^+ P^- = m_\pi^2$ to the
    contributions from the quark and gluon fields.
    (Right panel) divides out the total energy and plots the fractional contributions
    of the quark and gluon fields to the energy.
  }
  \label{fig:energy}
\end{figure}

With the variational solution, we can also decompose $m_\pi^2$ into contributions from the
quark and gluon fields.
This decomposition is presented in Fig.~\ref{fig:energy}.
The left panel plots the values of each contribution (in units of $g^2 N_c$),
both of which are positive and both of which vanish in the chiral limit.
The right panel divides out $m_\pi^2$ in order to obtain energy fractions.
Although the total amount of energy contributed by each field goes to zero in the chiral limit,
the energy fractions do not.
In fact, in the chiral limit---where the variational solution is the most trustworthy---the
energy fractions become $P_g^-/P^- \rightarrow 0.51$ and $P_q^-/P^- \rightarrow 0.49$.
For larger quark masses, the quarks unsurprisingly carry a greater fraction of the energy.

\begin{figure}
  \includegraphics[width=0.49\textwidth]{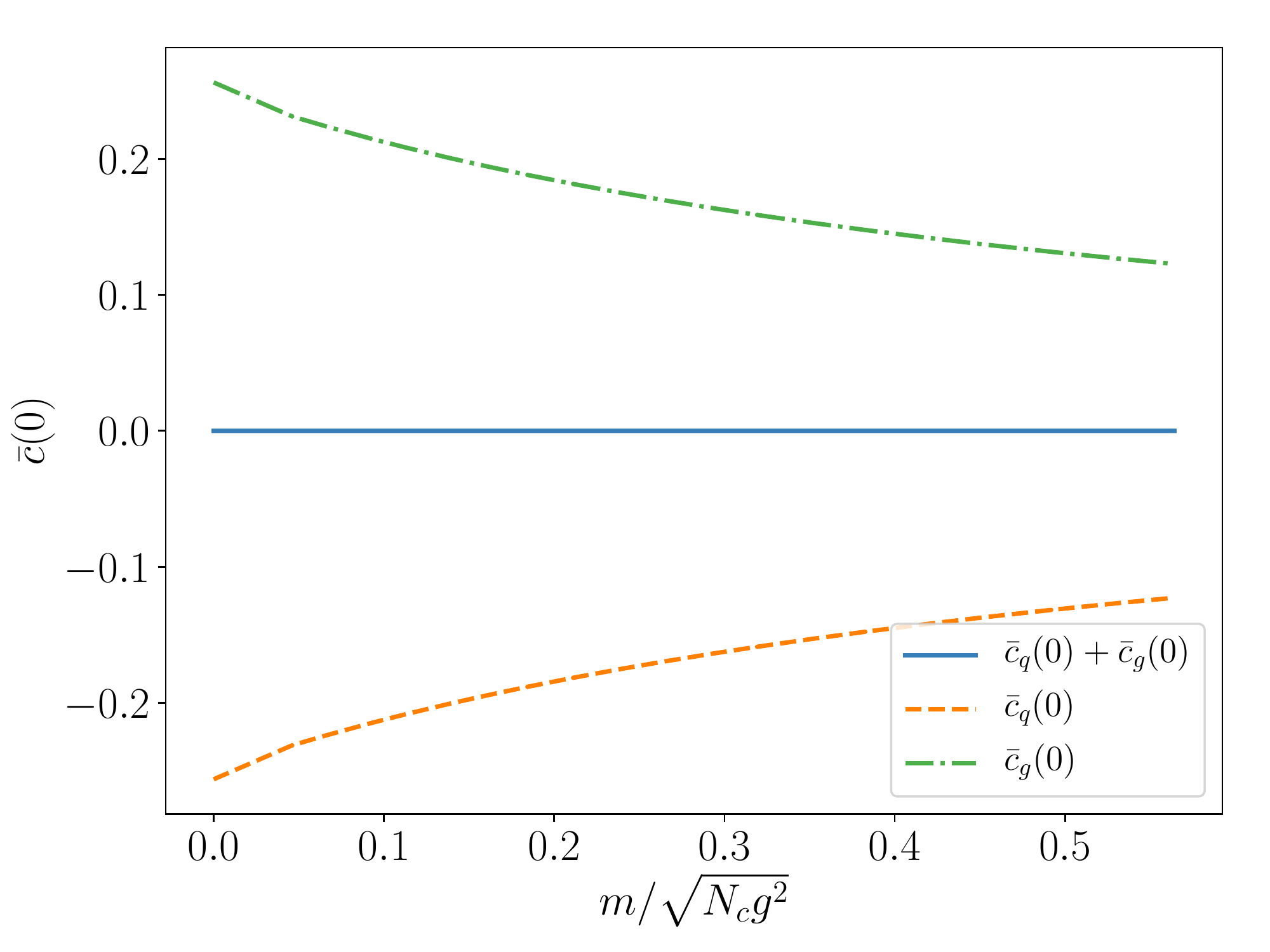}
  \caption{
    Plots of $\bar{c}_q(0)$ and $\bar{c}_g(0)$ as functions of quark mass,
    for the variational solution using Eq.~(\ref{eqn:ansatz}).
    Note that $\bar{c}_q(0) + \bar{c}_g(0) = 0$ as a consequence of energy conservation.
  }
  \label{fig:cbar}
\end{figure}

The energy fraction results also entail $\bar{c}_{q,g}(0)$ results through Eq.~(\ref{eqn:cbar}).
Results for these quantities as a function of quark mass are presented in Fig.~\ref{fig:cbar}.
It's worth noting that $\bar{c}_q(0) < 0$ and $\bar{c}_g(0) > 0$,
which is consistent with phenomenological estimates of $\bar{c}_q(0)$ for $(3+1)$-dimensional
QCD~\cite{Hatta:2018sqd,Lorce:2018egm}.
It is also worth noting that $\bar{c}_{q,g}(0)$ do not vanish in the chiral limit,
even though the total energy vanishes.

\begin{table}[t]
  \setlength{\tabcolsep}{0.5em}
  \renewcommand{\arraystretch}{1.3}
  \begin{tabular}{@{} cc|cccc @{}}
    \toprule
    $ m $ &
    $ m_\pi $ &
    $ \beta $ &
    $ \gamma $ &
    $ \mu^2 $ &
    $ g^2 N_c $ \\
    \hline
    $3.4$~MeV &
    $140$~MeV &
    $0.001183$ &
    $4.6 \cdot 10^{-6}$ &
    $0.007797$ &
    $7.897\cdot 10^{6}$~MeV$^2$ \\
    \bottomrule
  \end{tabular}
  \caption{
    't Hooft model parameters that reproduce a light quark mass of $3.4$~MeV
    and a pion mass of $140$~MeV.
  }
  \label{tab:parameters}
\end{table}

As a last interesting numerical consideration,
let us consider $m_\pi = 140$~MeV and $m = 3.4$~MeV.
Since $m_\pi$ is not linear in $m$ (see Eq.~(\ref{eqn:gmor}) for instance),
the $m_\pi/m$ ratio fixes all the parameters of the model.
The model parameters that reproduce this ratio are given in Tab.~\ref{tab:parameters}.
It is worth remarking that for the extremely small $\gamma$ (and $\beta$)
in this parameter set,
we are well within the range where the variational solution
using Eq.~(\ref{eqn:ansatz}) can be considered close to numerically exact.

\begin{table}[t]
  \setlength{\tabcolsep}{0.5em}
  \renewcommand{\arraystretch}{1.3}
  \begin{tabular}{@{} cccc @{}}
    \toprule
    $ \bar{c}_q(0) $ &
    $ \bar{c}_g(0) $ &
    $ P_q^- / P^- $ &
    $ P_g^- / P^- $ \\
    \hline
    $-0.294$ &
    $\phantom{-}0.294$ &
    $0.501$ &
    $0.499$ \\
    \bottomrule
  \end{tabular}
  \caption{
    Energy fractions and form factor $\bar{c}_{q,g}(0)$ values
    attributed to the quark and gluon fields inside the pion,
    using the model parameters in Tab.~\ref{tab:parameters}.
  }
  \label{tab:results}
\end{table}

With the model parameters fixed as in Tab.~\ref{tab:parameters},
energy fraction and $\bar{c}_{q,g}(0)$ values can be obtained.
The results are given in Tab.~\ref{tab:results}.
The gluon and quark energy fractions are nearly equal (half the total energy).

It it also interesting (though perhaps of limited instructional value)
to compare the $\bar{c}_q(0)$ result with $(3+1)$-dimensional QCD estimates.
Since $(3+1)$-dimensional QCD contains UV divergences that must be renormalized,
$\bar{c}_q(0)$ in this case can only be defined within a particular renormalization
scheme and at a particular renormalization scale.
Ref.~\cite{Hatta:2018sqd} gives $\bar{c}_q(\mu^2=\infty) = -0.103$ for two quark flavors.
The 't Hooft model $\bar{c}_q=-0.294$ is larger in magnitude,
but this may be an artifact of the number of dimensions.
However, the order of magnitude and negative sign are common between $D=1+1$ and $D=3+1$.


\section{Summary}
\label{sec:end}

The energy momentum tensor of the 't Hooft model is determined.
All the pion's light front momentum $P^+$ is carried by the quarks.
Although the gluon field is not dynamical,
it still carries a portion of the pion's light front energy $P^-$.
The pion energy can thus be decomposed into quark and gluon contributions,
which we have likewise obtained.
Results for a variety of current quark masses can be seen in
Fig.~\ref{fig:energy},
with numerical values for the empirical pion mass are given in
Tab.~\ref{tab:results}.

The model is solved using a variational method for small quark masses,
a case that has caused previous numerical difficulty.
We observe a GMOR-like relation and find that empirical values for the pion
and light quark masses fall within the domain where this solution is valid.
The quark and gluon fields each carry about equal amounts of the pion's energy.

\acknowledgments

We would like to thank Peter J.\ Ehlers for helpful discussions that helped contribute to this research.
This work was supported by the U.S. Department of Energy Office of Science, Office of Nuclear Physics under Award No. DE-FG02-97ER-41014.


\bibliography{millerOCT12}

\begin{thebibliography}{32}%
\makeatletter
\providecommand \@ifxundefined [1]{%
 \@ifx{#1\undefined}
}%
\providecommand \@ifnum [1]{%
 \ifnum #1\expandafter \@firstoftwo
 \else \expandafter \@secondoftwo
 \fi
}%
\providecommand \@ifx [1]{%
 \ifx #1\expandafter \@firstoftwo
 \else \expandafter \@secondoftwo
 \fi
}%
\providecommand \natexlab [1]{#1}%
\providecommand \enquote  [1]{``#1''}%
\providecommand \bibnamefont  [1]{#1}%
\providecommand \bibfnamefont [1]{#1}%
\providecommand \citenamefont [1]{#1}%
\providecommand \href@noop [0]{\@secondoftwo}%
\providecommand \href [0]{\begingroup \@sanitize@url \@href}%
\providecommand \@href[1]{\@@startlink{#1}\@@href}%
\providecommand \@@href[1]{\endgroup#1\@@endlink}%
\providecommand \@sanitize@url [0]{\catcode `\\12\catcode `\$12\catcode
  `\&12\catcode `\#12\catcode `\^12\catcode `\_12\catcode `\%12\relax}%
\providecommand \@@startlink[1]{}%
\providecommand \@@endlink[0]{}%
\providecommand \url  [0]{\begingroup\@sanitize@url \@url }%
\providecommand \@url [1]{\endgroup\@href {#1}{\urlprefix }}%
\providecommand \urlprefix  [0]{URL }%
\providecommand \Eprint [0]{\href }%
\providecommand \doibase [0]{http://dx.doi.org/}%
\providecommand \selectlanguage [0]{\@gobble}%
\providecommand \bibinfo  [0]{\@secondoftwo}%
\providecommand \bibfield  [0]{\@secondoftwo}%
\providecommand \translation [1]{[#1]}%
\providecommand \BibitemOpen [0]{}%
\providecommand \bibitemStop [0]{}%
\providecommand \bibitemNoStop [0]{.\EOS\space}%
\providecommand \EOS [0]{\spacefactor3000\relax}%
\providecommand \BibitemShut  [1]{\csname bibitem#1\endcsname}%
\let\auto@bib@innerbib\@empty
\bibitem [{\citenamefont {Ji}(1995)}]{Ji:1994av}%
  \BibitemOpen
  \bibfield  {author} {\bibinfo {author} {\bibfnamefont {X.-D.}\ \bibnamefont
  {Ji}},\ }\href {\doibase 10.1103/PhysRevLett.74.1071} {\bibfield  {journal}
  {\bibinfo  {journal} {Phys. Rev. Lett.}\ }\textbf {\bibinfo {volume} {74}},\
  \bibinfo {pages} {1071} (\bibinfo {year} {1995})},\ \Eprint
  {http://arxiv.org/abs/hep-ph/9410274} {arXiv:hep-ph/9410274} \BibitemShut
  {NoStop}%
\bibitem [{\citenamefont {He}\ and\ \citenamefont {Ji}(1995)}]{He:1994gz}%
  \BibitemOpen
  \bibfield  {author} {\bibinfo {author} {\bibfnamefont {H.-x.}\ \bibnamefont
  {He}}\ and\ \bibinfo {author} {\bibfnamefont {X.-D.}\ \bibnamefont {Ji}},\
  }\href {\doibase 10.1103/PhysRevD.52.2960} {\bibfield  {journal} {\bibinfo
  {journal} {Phys. Rev. D}\ }\textbf {\bibinfo {volume} {52}},\ \bibinfo
  {pages} {2960} (\bibinfo {year} {1995})},\ \Eprint
  {http://arxiv.org/abs/hep-ph/9412235} {arXiv:hep-ph/9412235} \BibitemShut
  {NoStop}%
\bibitem [{\citenamefont {Metz}\ \emph {et~al.}(2020)\citenamefont {Metz},
  \citenamefont {Pasquini},\ and\ \citenamefont {Rodini}}]{Metz:2020vxd}%
  \BibitemOpen
  \bibfield  {author} {\bibinfo {author} {\bibfnamefont {A.}~\bibnamefont
  {Metz}}, \bibinfo {author} {\bibfnamefont {B.}~\bibnamefont {Pasquini}}, \
  and\ \bibinfo {author} {\bibfnamefont {S.}~\bibnamefont {Rodini}},\ }\href
  {\doibase 10.1103/PhysRevD.102.114042} {\bibfield  {journal} {\bibinfo
  {journal} {Phys. Rev. D}\ }\textbf {\bibinfo {volume} {102}},\ \bibinfo
  {pages} {114042} (\bibinfo {year} {2020})},\ \Eprint
  {http://arxiv.org/abs/2006.11171} {arXiv:2006.11171 [hep-ph]} \BibitemShut
  {NoStop}%
\bibitem [{\citenamefont {Lorc\'e}\ \emph {et~al.}(2021)\citenamefont
  {Lorc\'e}, \citenamefont {Metz}, \citenamefont {Pasquini},\ and\
  \citenamefont {Rodini}}]{Lorce:2021xku}%
  \BibitemOpen
  \bibfield  {author} {\bibinfo {author} {\bibfnamefont {C.}~\bibnamefont
  {Lorc\'e}}, \bibinfo {author} {\bibfnamefont {A.}~\bibnamefont {Metz}},
  \bibinfo {author} {\bibfnamefont {B.}~\bibnamefont {Pasquini}}, \ and\
  \bibinfo {author} {\bibfnamefont {S.}~\bibnamefont {Rodini}},\ }\href
  {\doibase 10.1007/JHEP11(2021)121} {\bibfield  {journal} {\bibinfo  {journal}
  {JHEP}\ }\textbf {\bibinfo {volume} {11}},\ \bibinfo {pages} {121} (\bibinfo
  {year} {2021})},\ \Eprint {http://arxiv.org/abs/2109.11785} {arXiv:2109.11785
  [hep-ph]} \BibitemShut {NoStop}%
\bibitem [{\citenamefont {Ji}\ \emph {et~al.}(2021)\citenamefont {Ji},
  \citenamefont {Liu},\ and\ \citenamefont {Sch\"afer}}]{Ji:2021qgo}%
  \BibitemOpen
  \bibfield  {author} {\bibinfo {author} {\bibfnamefont {X.}~\bibnamefont
  {Ji}}, \bibinfo {author} {\bibfnamefont {Y.}~\bibnamefont {Liu}}, \ and\
  \bibinfo {author} {\bibfnamefont {A.}~\bibnamefont {Sch\"afer}},\ }\href
  {\doibase 10.1016/j.nuclphysb.2021.115537} {\bibfield  {journal} {\bibinfo
  {journal} {Nucl. Phys. B}\ }\textbf {\bibinfo {volume} {971}},\ \bibinfo
  {pages} {115537} (\bibinfo {year} {2021})},\ \Eprint
  {http://arxiv.org/abs/2105.03974} {arXiv:2105.03974 [hep-ph]} \BibitemShut
  {NoStop}%
\bibitem [{\citenamefont {Crewther}(1972)}]{Crewther:1972kn}%
  \BibitemOpen
  \bibfield  {author} {\bibinfo {author} {\bibfnamefont {R.~J.}\ \bibnamefont
  {Crewther}},\ }\href {\doibase 10.1103/PhysRevLett.28.1421} {\bibfield
  {journal} {\bibinfo  {journal} {Phys. Rev. Lett.}\ }\textbf {\bibinfo
  {volume} {28}},\ \bibinfo {pages} {1421} (\bibinfo {year}
  {1972})}\BibitemShut {NoStop}%
\bibitem [{\citenamefont {Chanowitz}\ and\ \citenamefont
  {Ellis}(1972)}]{Chanowitz:1972vd}%
  \BibitemOpen
  \bibfield  {author} {\bibinfo {author} {\bibfnamefont {M.~S.}\ \bibnamefont
  {Chanowitz}}\ and\ \bibinfo {author} {\bibfnamefont {J.~R.}\ \bibnamefont
  {Ellis}},\ }\href {\doibase 10.1016/0370-2693(72)90829-5} {\bibfield
  {journal} {\bibinfo  {journal} {Phys. Lett. B}\ }\textbf {\bibinfo {volume}
  {40}},\ \bibinfo {pages} {397} (\bibinfo {year} {1972})}\BibitemShut
  {NoStop}%
\bibitem [{\citenamefont {Collins}\ \emph {et~al.}(1977)\citenamefont
  {Collins}, \citenamefont {Duncan},\ and\ \citenamefont
  {Joglekar}}]{Collins:1976yq}%
  \BibitemOpen
  \bibfield  {author} {\bibinfo {author} {\bibfnamefont {J.~C.}\ \bibnamefont
  {Collins}}, \bibinfo {author} {\bibfnamefont {A.}~\bibnamefont {Duncan}}, \
  and\ \bibinfo {author} {\bibfnamefont {S.~D.}\ \bibnamefont {Joglekar}},\
  }\href {\doibase 10.1103/PhysRevD.16.438} {\bibfield  {journal} {\bibinfo
  {journal} {Phys. Rev. D}\ }\textbf {\bibinfo {volume} {16}},\ \bibinfo
  {pages} {438} (\bibinfo {year} {1977})}\BibitemShut {NoStop}%
\bibitem [{\citenamefont {Shifman}\ \emph {et~al.}(1978)\citenamefont
  {Shifman}, \citenamefont {Vainshtein},\ and\ \citenamefont
  {Zakharov}}]{Shifman:1978zn}%
  \BibitemOpen
  \bibfield  {author} {\bibinfo {author} {\bibfnamefont {M.~A.}\ \bibnamefont
  {Shifman}}, \bibinfo {author} {\bibfnamefont {A.~I.}\ \bibnamefont
  {Vainshtein}}, \ and\ \bibinfo {author} {\bibfnamefont {V.~I.}\ \bibnamefont
  {Zakharov}},\ }\href {\doibase 10.1016/0370-2693(78)90481-1} {\bibfield
  {journal} {\bibinfo  {journal} {Phys. Lett. B}\ }\textbf {\bibinfo {volume}
  {78}},\ \bibinfo {pages} {443} (\bibinfo {year} {1978})}\BibitemShut
  {NoStop}%
\bibitem [{\citenamefont {Dirac}(1949)}]{Dirac:1949cp}%
  \BibitemOpen
  \bibfield  {author} {\bibinfo {author} {\bibfnamefont {P.~A.~M.}\
  \bibnamefont {Dirac}},\ }\href {\doibase 10.1103/RevModPhys.21.392}
  {\bibfield  {journal} {\bibinfo  {journal} {Rev. Mod. Phys.}\ }\textbf
  {\bibinfo {volume} {21}},\ \bibinfo {pages} {392} (\bibinfo {year}
  {1949})}\BibitemShut {NoStop}%
\bibitem [{\citenamefont {Nambu}\ and\ \citenamefont
  {Jona-Lasinio}(1961{\natexlab{a}})}]{Nambu:1961tp}%
  \BibitemOpen
  \bibfield  {author} {\bibinfo {author} {\bibfnamefont {Y.}~\bibnamefont
  {Nambu}}\ and\ \bibinfo {author} {\bibfnamefont {G.}~\bibnamefont
  {Jona-Lasinio}},\ }\href {\doibase 10.1103/PhysRev.122.345} {\bibfield
  {journal} {\bibinfo  {journal} {Phys. Rev.}\ }\textbf {\bibinfo {volume}
  {122}},\ \bibinfo {pages} {345} (\bibinfo {year}
  {1961}{\natexlab{a}})}\BibitemShut {NoStop}%
\bibitem [{\citenamefont {Nambu}\ and\ \citenamefont
  {Jona-Lasinio}(1961{\natexlab{b}})}]{Nambu:1961fr}%
  \BibitemOpen
  \bibfield  {author} {\bibinfo {author} {\bibfnamefont {Y.}~\bibnamefont
  {Nambu}}\ and\ \bibinfo {author} {\bibfnamefont {G.}~\bibnamefont
  {Jona-Lasinio}},\ }\href {\doibase 10.1103/PhysRev.124.246} {\bibfield
  {journal} {\bibinfo  {journal} {Phys. Rev.}\ }\textbf {\bibinfo {volume}
  {124}},\ \bibinfo {pages} {246} (\bibinfo {year}
  {1961}{\natexlab{b}})}\BibitemShut {NoStop}%
\bibitem [{\citenamefont {Gell-Mann}\ \emph {et~al.}(1968)\citenamefont
  {Gell-Mann}, \citenamefont {Oakes},\ and\ \citenamefont
  {Renner}}]{Gell-Mann:1968hlm}%
  \BibitemOpen
  \bibfield  {author} {\bibinfo {author} {\bibfnamefont {M.}~\bibnamefont
  {Gell-Mann}}, \bibinfo {author} {\bibfnamefont {R.~J.}\ \bibnamefont
  {Oakes}}, \ and\ \bibinfo {author} {\bibfnamefont {B.}~\bibnamefont
  {Renner}},\ }\href {\doibase 10.1103/PhysRev.175.2195} {\bibfield  {journal}
  {\bibinfo  {journal} {Phys. Rev.}\ }\textbf {\bibinfo {volume} {175}},\
  \bibinfo {pages} {2195} (\bibinfo {year} {1968})}\BibitemShut {NoStop}%
\bibitem [{\citenamefont {Pagels}(1979)}]{Pagels:1978ba}%
  \BibitemOpen
  \bibfield  {author} {\bibinfo {author} {\bibfnamefont {H.}~\bibnamefont
  {Pagels}},\ }\href {\doibase 10.1103/PhysRevD.19.3080} {\bibfield  {journal}
  {\bibinfo  {journal} {Phys. Rev. D}\ }\textbf {\bibinfo {volume} {19}},\
  \bibinfo {pages} {3080} (\bibinfo {year} {1979})}\BibitemShut {NoStop}%
\bibitem [{\citenamefont {Chang}\ \emph {et~al.}(2013)\citenamefont {Chang},
  \citenamefont {Cloet}, \citenamefont {Cobos-Martinez}, \citenamefont
  {Roberts}, \citenamefont {Schmidt},\ and\ \citenamefont
  {Tandy}}]{Chang:2013pq}%
  \BibitemOpen
  \bibfield  {author} {\bibinfo {author} {\bibfnamefont {L.}~\bibnamefont
  {Chang}}, \bibinfo {author} {\bibfnamefont {I.~C.}\ \bibnamefont {Cloet}},
  \bibinfo {author} {\bibfnamefont {J.~J.}\ \bibnamefont {Cobos-Martinez}},
  \bibinfo {author} {\bibfnamefont {C.~D.}\ \bibnamefont {Roberts}}, \bibinfo
  {author} {\bibfnamefont {S.~M.}\ \bibnamefont {Schmidt}}, \ and\ \bibinfo
  {author} {\bibfnamefont {P.~C.}\ \bibnamefont {Tandy}},\ }\href {\doibase
  10.1103/PhysRevLett.110.132001} {\bibfield  {journal} {\bibinfo  {journal}
  {Phys. Rev. Lett.}\ }\textbf {\bibinfo {volume} {110}},\ \bibinfo {pages}
  {132001} (\bibinfo {year} {2013})},\ \Eprint {http://arxiv.org/abs/1301.0324}
  {arXiv:1301.0324 [nucl-th]} \BibitemShut {NoStop}%
\bibitem [{\citenamefont {'t~Hooft}(1974)}]{tHooft:1974pnl}%
  \BibitemOpen
  \bibfield  {author} {\bibinfo {author} {\bibfnamefont {G.}~\bibnamefont
  {'t~Hooft}},\ }\href {\doibase 10.1016/0550-3213(74)90088-1} {\bibfield
  {journal} {\bibinfo  {journal} {Nucl. Phys. B}\ }\textbf {\bibinfo {volume}
  {75}},\ \bibinfo {pages} {461} (\bibinfo {year} {1974})}\BibitemShut
  {NoStop}%
\bibitem [{\citenamefont {Ellis}(1977)}]{Ellis:1977mw}%
  \BibitemOpen
  \bibfield  {author} {\bibinfo {author} {\bibfnamefont {J.~R.}\ \bibnamefont
  {Ellis}},\ }\href@noop {} {\bibfield  {journal} {\bibinfo  {journal} {Acta
  Phys. Polon. B}\ }\textbf {\bibinfo {volume} {8}},\ \bibinfo {pages} {1019}
  (\bibinfo {year} {1977})}\BibitemShut {NoStop}%
\bibitem [{\citenamefont {De~T\'eramond}\ and\ \citenamefont
  {Brodsky}(2021)}]{DeTeramond:2021jnn}%
  \BibitemOpen
  \bibfield  {author} {\bibinfo {author} {\bibfnamefont {G.~F.}\ \bibnamefont
  {De~T\'eramond}}\ and\ \bibinfo {author} {\bibfnamefont {S.~J.}\ \bibnamefont
  {Brodsky}},\ }\href@noop {} {\  (\bibinfo {year} {2021})},\ \Eprint
  {http://arxiv.org/abs/2103.10950} {arXiv:2103.10950 [hep-ph]} \BibitemShut
  {NoStop}%
\bibitem [{\citenamefont {Li}\ and\ \citenamefont {Vary}(2021)}]{Li:2021jqb}%
  \BibitemOpen
  \bibfield  {author} {\bibinfo {author} {\bibfnamefont {Y.}~\bibnamefont
  {Li}}\ and\ \bibinfo {author} {\bibfnamefont {J.~P.}\ \bibnamefont {Vary}},\
  }\href@noop {} {\  (\bibinfo {year} {2021})},\ \Eprint
  {http://arxiv.org/abs/2103.09993} {arXiv:2103.09993 [hep-ph]} \BibitemShut
  {NoStop}%
\bibitem [{\citenamefont {Ahmady}\ \emph
  {et~al.}(2021{\natexlab{a}})\citenamefont {Ahmady}, \citenamefont {Dahiya},
  \citenamefont {Kaur}, \citenamefont {Mondal}, \citenamefont {Sandapen},\ and\
  \citenamefont {Sharma}}]{Ahmady:2021lsh}%
  \BibitemOpen
  \bibfield  {author} {\bibinfo {author} {\bibfnamefont {M.}~\bibnamefont
  {Ahmady}}, \bibinfo {author} {\bibfnamefont {H.}~\bibnamefont {Dahiya}},
  \bibinfo {author} {\bibfnamefont {S.}~\bibnamefont {Kaur}}, \bibinfo {author}
  {\bibfnamefont {C.}~\bibnamefont {Mondal}}, \bibinfo {author} {\bibfnamefont
  {R.}~\bibnamefont {Sandapen}}, \ and\ \bibinfo {author} {\bibfnamefont
  {N.}~\bibnamefont {Sharma}},\ }\href {\doibase
  10.1016/j.physletb.2021.136754} {\bibfield  {journal} {\bibinfo  {journal}
  {Phys. Lett. B}\ }\textbf {\bibinfo {volume} {823}},\ \bibinfo {pages}
  {136754} (\bibinfo {year} {2021}{\natexlab{a}})},\ \Eprint
  {http://arxiv.org/abs/2105.01018} {arXiv:2105.01018 [hep-ph]} \BibitemShut
  {NoStop}%
\bibitem [{\citenamefont {Ahmady}\ \emph
  {et~al.}(2021{\natexlab{b}})\citenamefont {Ahmady}, \citenamefont {Kaur},
  \citenamefont {MacKay}, \citenamefont {Mondal},\ and\ \citenamefont
  {Sandapen}}]{Ahmady:2021yzh}%
  \BibitemOpen
  \bibfield  {author} {\bibinfo {author} {\bibfnamefont {M.}~\bibnamefont
  {Ahmady}}, \bibinfo {author} {\bibfnamefont {S.}~\bibnamefont {Kaur}},
  \bibinfo {author} {\bibfnamefont {S.~L.}\ \bibnamefont {MacKay}}, \bibinfo
  {author} {\bibfnamefont {C.}~\bibnamefont {Mondal}}, \ and\ \bibinfo {author}
  {\bibfnamefont {R.}~\bibnamefont {Sandapen}},\ }\href {\doibase
  10.1103/PhysRevD.104.074013} {\bibfield  {journal} {\bibinfo  {journal}
  {Phys. Rev. D}\ }\textbf {\bibinfo {volume} {104}},\ \bibinfo {pages}
  {074013} (\bibinfo {year} {2021}{\natexlab{b}})},\ \Eprint
  {http://arxiv.org/abs/2108.03482} {arXiv:2108.03482 [hep-ph]} \BibitemShut
  {NoStop}%
\bibitem [{\citenamefont {Brodsky}\ \emph {et~al.}(2015)\citenamefont
  {Brodsky}, \citenamefont {de~Teramond}, \citenamefont {Dosch},\ and\
  \citenamefont {Erlich}}]{Brodsky:2014yha}%
  \BibitemOpen
  \bibfield  {author} {\bibinfo {author} {\bibfnamefont {S.~J.}\ \bibnamefont
  {Brodsky}}, \bibinfo {author} {\bibfnamefont {G.~F.}\ \bibnamefont
  {de~Teramond}}, \bibinfo {author} {\bibfnamefont {H.~G.}\ \bibnamefont
  {Dosch}}, \ and\ \bibinfo {author} {\bibfnamefont {J.}~\bibnamefont
  {Erlich}},\ }\href {\doibase 10.1016/j.physrep.2015.05.001} {\bibfield
  {journal} {\bibinfo  {journal} {Phys. Rept.}\ }\textbf {\bibinfo {volume}
  {584}},\ \bibinfo {pages} {1} (\bibinfo {year} {2015})},\ \Eprint
  {http://arxiv.org/abs/1407.8131} {arXiv:1407.8131 [hep-ph]} \BibitemShut
  {NoStop}%
\bibitem [{\citenamefont {Sheckler}\ and\ \citenamefont
  {Miller}(2021)}]{Sheckler:2020fbt}%
  \BibitemOpen
  \bibfield  {author} {\bibinfo {author} {\bibfnamefont {A.~B.}\ \bibnamefont
  {Sheckler}}\ and\ \bibinfo {author} {\bibfnamefont {G.~A.}\ \bibnamefont
  {Miller}},\ }\href {\doibase 10.1103/PhysRevD.103.096018} {\bibfield
  {journal} {\bibinfo  {journal} {Phys. Rev. D}\ }\textbf {\bibinfo {volume}
  {103}},\ \bibinfo {pages} {096018} (\bibinfo {year} {2021})},\ \Eprint
  {http://arxiv.org/abs/2101.00100} {arXiv:2101.00100 [hep-ph]} \BibitemShut
  {NoStop}%
\bibitem [{\citenamefont {Chabysheva}\ and\ \citenamefont
  {Hiller}(2013)}]{Chabysheva:2012fe}%
  \BibitemOpen
  \bibfield  {author} {\bibinfo {author} {\bibfnamefont {S.~S.}\ \bibnamefont
  {Chabysheva}}\ and\ \bibinfo {author} {\bibfnamefont {J.~R.}\ \bibnamefont
  {Hiller}},\ }\href {\doibase 10.1016/j.aop.2013.06.016} {\bibfield  {journal}
  {\bibinfo  {journal} {Annals Phys.}\ }\textbf {\bibinfo {volume} {337}},\
  \bibinfo {pages} {143} (\bibinfo {year} {2013})},\ \Eprint
  {http://arxiv.org/abs/1207.7128} {arXiv:1207.7128 [hep-ph]} \BibitemShut
  {NoStop}%
\bibitem [{\citenamefont {Brower}\ \emph {et~al.}(1979)\citenamefont {Brower},
  \citenamefont {Spence},\ and\ \citenamefont {Weis}}]{PhysRevD.19.3024}%
  \BibitemOpen
  \bibfield  {author} {\bibinfo {author} {\bibfnamefont {R.~C.}\ \bibnamefont
  {Brower}}, \bibinfo {author} {\bibfnamefont {W.~L.}\ \bibnamefont {Spence}},
  \ and\ \bibinfo {author} {\bibfnamefont {J.~H.}\ \bibnamefont {Weis}},\
  }\href {\doibase 10.1103/PhysRevD.19.3024} {\bibfield  {journal} {\bibinfo
  {journal} {Phys. Rev. D}\ }\textbf {\bibinfo {volume} {19}},\ \bibinfo
  {pages} {3024} (\bibinfo {year} {1979})}\BibitemShut {NoStop}%
\bibitem [{\citenamefont {Kugo}\ and\ \citenamefont
  {Ojima}(1979)}]{Kugo:1979gm}%
  \BibitemOpen
  \bibfield  {author} {\bibinfo {author} {\bibfnamefont {T.}~\bibnamefont
  {Kugo}}\ and\ \bibinfo {author} {\bibfnamefont {I.}~\bibnamefont {Ojima}},\
  }\href {\doibase 10.1143/PTPS.66.1} {\bibfield  {journal} {\bibinfo
  {journal} {Prog. Theor. Phys. Suppl.}\ }\textbf {\bibinfo {volume} {66}},\
  \bibinfo {pages} {1} (\bibinfo {year} {1979})}\BibitemShut {NoStop}%
\bibitem [{\citenamefont {Leader}\ and\ \citenamefont
  {Lorc\'e}(2014)}]{Leader:2013jra}%
  \BibitemOpen
  \bibfield  {author} {\bibinfo {author} {\bibfnamefont {E.}~\bibnamefont
  {Leader}}\ and\ \bibinfo {author} {\bibfnamefont {C.}~\bibnamefont
  {Lorc\'e}},\ }\href {\doibase 10.1016/j.physrep.2014.02.010} {\bibfield
  {journal} {\bibinfo  {journal} {Phys. Rept.}\ }\textbf {\bibinfo {volume}
  {541}},\ \bibinfo {pages} {163} (\bibinfo {year} {2014})},\ \Eprint
  {http://arxiv.org/abs/1309.4235} {arXiv:1309.4235 [hep-ph]} \BibitemShut
  {NoStop}%
\bibitem [{\citenamefont {Ehlers}(2022)}]{Ehlers:2022oal}%
  \BibitemOpen
  \bibfield  {author} {\bibinfo {author} {\bibfnamefont {P.~J.}\ \bibnamefont
  {Ehlers}},\ }\href@noop {} {\  (\bibinfo {year} {2022})},\ \Eprint
  {http://arxiv.org/abs/2209.09867} {arXiv:2209.09867 [hep-ph]} \BibitemShut
  {NoStop}%
\bibitem [{\citenamefont {Gribov}(1978)}]{Gribov:1977wm}%
  \BibitemOpen
  \bibfield  {author} {\bibinfo {author} {\bibfnamefont {V.~N.}\ \bibnamefont
  {Gribov}},\ }\href {\doibase 10.1016/0550-3213(78)90175-X} {\bibfield
  {journal} {\bibinfo  {journal} {Nucl. Phys. B}\ }\textbf {\bibinfo {volume}
  {139}},\ \bibinfo {pages} {1} (\bibinfo {year} {1978})}\BibitemShut {NoStop}%
\bibitem [{{\relax DLMF}()}]{NIST:DLMF}%
  \BibitemOpen
  {\relax DLMF},\ \href {http://dlmf.nist.gov/} {\enquote {\bibinfo {title}
  {{\it NIST Digital Library of Mathematical Functions}},}\ }\bibinfo
  {howpublished} {http://dlmf.nist.gov/, Release 1.1.0 of 2020-12-15},\
  \bibinfo {note} {f.~W.~J. Olver, A.~B. {Olde Daalhuis}, D.~W. Lozier, B.~I.
  Schneider, R.~F. Boisvert, C.~W. Clark, B.~R. Miller, B.~V. Saunders, H.~S.
  Cohl, and M.~A. McClain, eds.}\BibitemShut {Stop}%
\bibitem [{\citenamefont {Hatta}\ \emph {et~al.}(2018)\citenamefont {Hatta},
  \citenamefont {Rajan},\ and\ \citenamefont {Tanaka}}]{Hatta:2018sqd}%
  \BibitemOpen
  \bibfield  {author} {\bibinfo {author} {\bibfnamefont {Y.}~\bibnamefont
  {Hatta}}, \bibinfo {author} {\bibfnamefont {A.}~\bibnamefont {Rajan}}, \ and\
  \bibinfo {author} {\bibfnamefont {K.}~\bibnamefont {Tanaka}},\ }\href
  {\doibase 10.1007/JHEP12(2018)008} {\bibfield  {journal} {\bibinfo  {journal}
  {JHEP}\ }\textbf {\bibinfo {volume} {12}},\ \bibinfo {pages} {008} (\bibinfo
  {year} {2018})},\ \Eprint {http://arxiv.org/abs/1810.05116} {arXiv:1810.05116
  [hep-ph]} \BibitemShut {NoStop}%
\bibitem [{\citenamefont {Lorc\'e}\ \emph {et~al.}(2019)\citenamefont
  {Lorc\'e}, \citenamefont {Moutarde},\ and\ \citenamefont
  {Trawi\'nski}}]{Lorce:2018egm}%
  \BibitemOpen
  \bibfield  {author} {\bibinfo {author} {\bibfnamefont {C.}~\bibnamefont
  {Lorc\'e}}, \bibinfo {author} {\bibfnamefont {H.}~\bibnamefont {Moutarde}}, \
  and\ \bibinfo {author} {\bibfnamefont {A.~P.}\ \bibnamefont {Trawi\'nski}},\
  }\href {\doibase 10.1140/epjc/s10052-019-6572-3} {\bibfield  {journal}
  {\bibinfo  {journal} {Eur. Phys. J. C}\ }\textbf {\bibinfo {volume} {79}},\
  \bibinfo {pages} {89} (\bibinfo {year} {2019})},\ \Eprint
  {http://arxiv.org/abs/1810.09837} {arXiv:1810.09837 [hep-ph]} \BibitemShut
  {NoStop}%
\end{thebibliography}%

\end{document}